\documentclass[conference]{IEEEtran}
\IEEEoverridecommandlockouts
\usepackage{amsmath}
\usepackage{amsfonts}
\usepackage{algorithm}
\usepackage{algorithmic}
\usepackage{algorithm}
\usepackage{array}
\usepackage{textcomp}
\usepackage{stfloats}
\usepackage{url}
\usepackage{verbatim}
\usepackage{graphicx}
\usepackage{cite}
\usepackage{enumitem}
\usepackage{amsthm}
\usepackage[T1]{fontenc}
\usepackage{amssymb}
\usepackage{epsfig}
\usepackage{psfrag}
\usepackage{latexsym}
\usepackage{lettrine}
\usepackage{color}
\usepackage{multirow}
\usepackage{subfigure}
\usepackage{enumitem}
\usepackage{bm}
\usepackage{array}
\usepackage{hyperref}					
\usepackage{balance}
\hypersetup{colorlinks,
	linkcolor=blue,%
	anchorcolor=blue,
    citecolor=blue}

\usepackage{balance}
\usepackage{cite}
\usepackage{amsmath,amssymb,amsfonts}
\usepackage{algorithmic}
\usepackage{graphicx}
\usepackage{textcomp}
\usepackage{xcolor}
\def\BibTeX{{\rm B\kern-.05em{\sc i\kern-.025em b}\kern-.08em
    T\kern-.1667em\lower.7ex\hbox{E}\kern-.125emX}}
\begin{document}

\makeatletter
\newcommand{\linebreakand}{%
  \end{@IEEEauthorhalign}
  \hfill\mbox{}\par
  \mbox{}\hfill\begin{@IEEEauthorhalign}
}
\makeatother

\title{ SNR-Dependent Mismatched Filtering for Bistatic OFDM Ranging\\
\thanks{This work was supported in part by the Mobile Information Networks-National Science and Technology Major Project under Grant 2025ZD1302000, and in part by the National Natural Science Foundation of China (NSFC) under Grant 62522107 and Grant 62331023.}
}

\author{
\IEEEauthorblockN{Ying Zhang\IEEEauthorrefmark{1},
Fan Liu\IEEEauthorrefmark{2},
Yifeng Xiong\IEEEauthorrefmark{3},
Jie Yang\IEEEauthorrefmark{2},
Xinyi Wang\IEEEauthorrefmark{4}, and
Shi Jin\IEEEauthorrefmark{2}}
\IEEEauthorblockA{\IEEEauthorrefmark{1}Southern University of Science and Technology, Shenzhen, China}
\IEEEauthorblockA{\IEEEauthorrefmark{2}Southeast University, Nanjing, China}
\IEEEauthorblockA{\IEEEauthorrefmark{3}Beijing University of Posts and Telecommunications, Beijing, China}
\IEEEauthorblockA{\IEEEauthorrefmark{4}Beijing Institute of Technology, Beijing, China}
\IEEEauthorblockA{Emails: zhangying2024@mail.sustech.edu.cn, fan.liu@seu.edu.cn, yifengxiong@bupt.edu.cn, \\yangjie@seu.edu.cn, wangxinyi@bit.edu.cn, jinshi@seu.edu.cn}
}

\maketitle

\begin{abstract}
This paper investigates the ranging performance of a bistatic integrated sensing and communications (ISAC) system employing orthogonal frequency-division multiplexing (OFDM), in which an ISAC transmitter emits a communication waveform carrying random data symbols, and a separate receiver performs ranging by correlating the received signal with a locally demodulated symbol sequence. Owing to inevitable demodulation errors, the ranging processor operates under mismatched filtering rather than ideal matched filtering, resulting in a delay-domain correlation response whose sidelobe structure explicitly depends on the signal-to-noise ratio (SNR). Focusing on frequency-flat fading channels, we derive closed-form expressions for the expected sidelobe level (ESL) and the average mainlobe level of the resulting mismatched ranging response for BPSK, QPSK, and general square QAM constellations. The analysis quantitatively characterizes how SNR-driven symbol decision errors reshape the delay-domain sidelobe behavior, thereby providing analytical insight into the SNR-dependent scaling behavior of ranging performance in bistatic OFDM-based ISAC systems. Simulation results validate the theoretical derivations and confirm the accuracy of the proposed analysis.
\end{abstract}


\section{Introduction}
High-precision localization and high-rate data transmission are fundamental to emerging applications such as Smart Cities and Homes,
Vehicular Networks, and the Low-Altitude Economy\cite{10041914, 11206742}. In conventional systems, sensing and communication are typically implemented by separate modules, leading to duplicated front-end hardware and frame structures, inefficient spectrum utilization, and mutual interference. These limitations have motivated growing interest in integrated sensing and communication (ISAC) technologies \cite{9540344, 10833779}.

Depending on waveform design strategies, ISAC systems can be broadly classified into three categories: communication-centric, sensing-centric, and joint designs \cite{7855671}. Compatible with existing cellular protocols, communication-centric designs can be readily integrated into current network infrastructures with minimal hardware modifications while preserving communication performance. In existing communication-centric ISAC systems, sensing is mainly performed using reference signals that occupy only a small portion of the time-frequency resources, which motivates the development of sensing techniques that exploit random data signals occupying most of the available resources\cite{10904093, 10969844, 11322556}. Several studies have examined the sensing performance of random communication data payload signals \cite{11391499, 10645253, zhang2026discreteambiguityfunctionsrandom}. However, existing studies on random communication signals have mainly focused on monostatic ISAC scenarios where the data symbols are perfectly known. In contrast, their sensing performance in bistatic scenarios remains largely unexplored. 
 
 To address this gap, this work investigates the ranging performance of a bistatic OFDM-based ISAC system, where the transmitter sends communication waveforms carrying random data symbols and a separate receiver performs ranging by correlating the sensing-link observations with a locally demodulated symbol sequence. Unlike the ideal matched filtering assumption commonly employed in monostatic ISAC, the considered ranging process is affected by symbol decision errors, resulting in mismatched filtering. 
 
 In this work, we focus on the frequency-flat fading case (i.e., single-target scenario) and derive closed-form expressions for the expected sidelobe level (ESL) and average mainlobe level under BPSK, QPSK, and general square QAM modulations. The proposed analysis reveals the signal-to-noise ratio (SNR) dependent scaling behavior of the mismatched ranging response and is validated through numerical simulations.

{\emph{Notations}}: Matrices are denoted by bold uppercase (e.g., $\mathbf{U}$), vectors by bold lowercase (e.g., $\mathbf{x}$), and scalars by normal font (e.g., $N$). The $n$th entry of $\mathbf{s}$ is $s_n$. $\circledast$ and $\odot$ denote the circular convolution and Hadamard product. $\left(\cdot\right)^T$, $\left(\cdot\right)^H$, and $\left(\cdot\right)^*$ denote transpose, Hermitian transpose, and complex conjugate. $\mathbb{E}(\cdot)$ denotes expectation.

\section{System Model}
\subsection{ISAC Signal Model}
Consider an OFDM-based bistatic ISAC system, where the transmitter emits communication symbols that are also exploited for ranging at a separate receiver. Let $\mathbf{s} = [s_1, s_2, ..., s_N]^T \in \mathbb{C}^N$
denote the vector of $N$ transmitted communication symbols. After OFDM modulation, the corresponding discrete-time transmit signal can be expressed as
\begin{align}
    \mathbf{x} = \mathbf{F}_N^H \mathbf{s},
\end{align}
where \(\mathbf{F}_N \in \mathbb{C}^{N\times N}\) denotes the normalized discrete Fourier transform (DFT) matrix.

The time-domain impulse response of the considered channel is modeled as $\mathbf{h} = \left[h[0], h[1], ..., h[N-1] \right]^T$, where
\begin{align}
    h[n] = \mu \delta[n-\tau],\quad n = 0,1,...,N-1,
\end{align}
with $\mu \in \mathbb{C}$ denoting the complex channel coefficient and $\tau$ representing the corresponding discrete delay index. Specifically, the coefficient $\mu$ affects both the amplitude and phase of the received signal, while $\tau$ specifies the target delay in the discrete-time domain, thereby indicating the location of the mainlobe in the delay profile. Here, $\delta[\cdot]$ denotes the discrete-time Kronecker delta function. The corresponding frequency-domain channel response is given by
\begin{align}
    \tilde{\mathbf{h}} = \mu \left[1, e^{-j \frac{2 \pi}{N} \tau}, ..., e^{-j \frac{2 \pi}{N} \tau(N-1)} \right]^T.
\end{align}
At the receiver side, the echo signal is received through two sets of antennas and RF chains, one for communication processing and the other for sensing processing, which are given by 
\begin{align}
    \mathbf{y}_1 = \mathbf{h} \circledast  \mathbf{x} + \mathbf{z}_1, \quad \mathbf{y}_2 = \mathbf{h} \circledast  \mathbf{x} + \mathbf{z}_2,
\end{align}
where \(\mathbf{y}_1\) and \(\mathbf{y}_2\) represent the received signals associated with the communication link and sensing link, respectively, and $\mathbf{z}_1$ and $\mathbf{z}_2$ are two independent additive white Gaussian noise (AWGN) vectors. Under the normalized symbol energy assumption, i.e., $E_s=1$, the SNR is defined as
\begin{align}
    \gamma := {E_s}/{N_0} = {1}/{N_0},
\end{align}
where \(N_0\) denotes the noise power spectral density.

\subsection{Sensing Signal Processing and Performance Metric}
After OFDM demodulation and hard-decision detection based on the received signal over the communication link, the detected symbol vector is obtained as
\begin{align}
    \hat{\mathbf{s}} = [\hat{s}_1, \hat{s}_2,...,\hat{s}_N]^T.
\end{align}
For the sensing link, the received signal is transformed into the frequency-domain as
\begin{align}
    \tilde{\mathbf{y}}_2 = \mathbf{F}_N \mathbf{y}_2 = \tilde{\mathbf{h}} \odot \mathbf{s} + \tilde{\mathbf{z}}_{2},
\end{align}
where $\tilde{\mathbf{z}}_{2}$ denotes the frequency-domain noise vector. 

The detected symbols $\hat{\mathbf{s}}$ from the communication link are then employed as the reference sequence to perform frequency-domain filtering on the received sensing signal. The corresponding frequency-domain output is given by
\begin{align} \label{MF in frequency domain}
    \mathbf{r}_f = \tilde{\mathbf{y}}_2 \odot \hat{\mathbf{s}}^* 
    & = \tilde{\mathbf{h}} \odot \mathbf{c} + \tilde{\mathbf{z}}_{2} \odot \hat{\mathbf{s}}^*,
\end{align}
where
\begin{align}
    \mathbf{c}  := \mathbf{s} \odot \hat{\mathbf{s}}^*  = [c_1, c_2, ..., c_N]^T 
    = [s_1 \hat{s}_1^*, s_2 \hat{s}_2^*, ..., s_N \hat{s}_N^*]^T.
\end{align}
Applying the inverse discrete Fourier transform (IDFT) to \eqref{MF in frequency domain} yields the time-domain correlation response as
\begin{align}
    \mathbf{r} 
    = \sqrt{N} \mathbf{F}_N^H \left( \tilde{\mathbf{h}} \odot \mathbf{c} + \tilde{\mathbf{z}}_{2} \odot \hat{\mathbf{s}}^* \right).
\end{align}
Accordingly, the $k$-th ($k = 0,1, ..., N-1$) element of $\mathbf{r}$ is
\begin{align}\label{MF time domain 1}
    r_k 
    &\nonumber = \sqrt{N} \mathbf{f}_{k+1}^H \left( \tilde{\mathbf{h}} \odot \mathbf{c} + \tilde{\mathbf{z}}_{2} \odot \hat{\mathbf{s}}^* \right)\\
    & = \mu \sum_{n=1}^{N}  c_n e^{j \frac{2\pi}{N} (k-\tau)(n-1)} + \sum_{n=1}^{N} \tilde{z}_{2,n} \hat{s}_n^* e^{j \frac{2\pi}{N} k(n-1)},
\end{align}
where $\mathbf{f}_k$ is the $k$-th column of $\mathbf{F}_N$.

To facilitate the subsequent analysis, the channel delay index is set to $\tau=0$, and the corresponding channel coefficient is set to $\mu = 1$ throughout this work, corresponding to a single target at the zero-delay bin. The extension to the non-zero-delay case is straightforward, because a non-zero delay only shifts the mainlobe from $k=0$ to $k=\tau$, whereas the sidelobe analysis is unaffected. Since the transmitted signal is random, the resulting correlation response is also random. Accordingly, the expected sidelobe level (ESL) is adopted as the performance metric and can be expressed as
\begin{align}\label{ESL}
    &\nonumber\mathbb{E}(|r_k|^2) 
     = \mathbb{E}\left(  \left|  \sum_{n=1}^{N} c_n e^{j \frac{2\pi}{N} k(n-1)} + \sum_{n=1}^{N} \tilde{z}_{2,n} \hat{s}_n^* e^{j \frac{2\pi}{N} k(n-1)}\right|^2 \right) \\
    & =   \mathbb{E}\left(  \left| \sum_{n=1}^N c_n e^{j \frac{2 \pi}{N} k (n-1)} \right|^2 \right) + \mathbb{E}\left(  \left| \sum_{n=1}^N \tilde{z}_{2,n} \hat{s}_n^* e^{j \frac{2 \pi}{N} k (n-1)}\right|^2 \right),
\end{align}
where the expectation is with respect to the random symbols and noise. Note that the cross term vanishes due to the independence of $\mathbf{z}_2$ and $\hat{\mathbf{s}}$. Moreover, $\mathbb{E}(|r_0|^2)$ denotes the average mainlobe level, and $\mathbb{E}(|r_k|^2), k \ne 0$ characterizes the ESL at the index $k$.

We first consider the signal-dependent term in \eqref{ESL}, which can be expanded as
\begin{align} \label{The first term of ESL}
    &\nonumber \mathbb{E} \left( \left| \sum_{n=1}^N c_n e^{j \frac{2 \pi}{N} k (n-1)} \right|^2 \right)  = \mathbb{E} \left( \sum_{n=1}^N \sum_{m=1}^N c_n c_m^* e^{j \frac{2 \pi}{N} k (n-m)} \right) \\
    &\nonumber = \mathbb{E} \left( \sum_{n=1}^N |c_n|^2 + \sum_{n=1}^N \sum_{m=1, m \ne n}^N c_n c_m^* e^{j \frac{2 \pi}{N} k (n-m)}\right) \\
    &\nonumber = N \mathbb{E}(|c_n|^2) + \left| \mathbb{E}(c_n) \right|^2 \sum_{n=1}^N \sum_{m=1, m \ne n}^Ne^{j \frac{2 \pi}{N} k (n-m)} \\
    &:= N M_2 + |M_1|^2 \sum_{n=1}^N \sum_{m=1, m \ne n}^Ne^{j \frac{2 \pi}{N} k (n-m)},
\end{align}
where
\begin{align}
    & M_2 = \mathbb{E}(|c_n|^2) = \mathbb{E}(|s_n|^2 |\hat{s}_n|^2), \\
    & M_1 =  \mathbb{E}(c_n)  = \mathbb{E}(s_n \hat{s}_n^*).
\end{align}
Therefore, the first term of \eqref{ESL} can be simplified as
\begin{align} \label{T1}
    T_1 = \left\{ 
    \begin{aligned}
        &  N M_2 + N(N-1) |M_1|^2, \quad k = 0,\\
        &  N (M_2 - |M_1|^2), \quad k \ne 0.
    \end{aligned}
    \right.    
\end{align}
Next, for the noise-dependent term in \eqref{ESL}, we have
\begin{align} \label{T2}
    T_2 = \mathbb{E}\left(  \left| \sum_{n=1}^N \tilde{z}_{2,n} \hat{s}_n^* e^{j \frac{2 \pi}{N} k (n-1)}\right|^2 \right) = N_0 N \mathbb{E}(|\hat{s}_n|^2).
\end{align}
By combining the above two terms, the average mainlobe level and ESL are obtained as
\begin{align} \label{ESL 2}
    & \nonumber \mathbb{E}(|r_k|^2) = \\
    & \left\{ 
    \begin{aligned}
        &  N M_2 + N(N-1) |M_1|^2 + N_0 N \mathbb{E}(|\hat{s}_n|^2), \quad k = 0,\\
        &  N (M_2 - |M_1|^2) + N_0 N \mathbb{E}(|\hat{s}_n|^2), \quad k \ne 0.
    \end{aligned}
    \right.    
\end{align}

\section{Statistical Characterization of the Mismatched Filtering}
In this section, the ESL of the mismatched-filter response is characterized through the following case studies.

\subsection{BPSK}
We first consider the BPSK case, where each transmitted symbol is drawn from the binary constellation $\{-1,1\}$, and thus $M_2$ is directly given by
\begin{align}
    M_2 = \mathbb{E}(|s_n|^2 |\hat{s}_n|^2) = 1.
\end{align}
Under AWGN and hard-decision detection, the conditional decision probabilities are given by
\begin{align}
\begin{aligned}
    & \mathrm{Pr}(c_n = 1) = 1 - Q(\sqrt{2 \gamma}), \\
    & \mathrm{Pr}(c_n = -1) = Q(\sqrt{2 \gamma}), \\
\end{aligned}
\end{align}
where $Q(\cdot)$ is the Q-function. Based on the above probabilities, $M_1$ can be calculated as
\begin{align}
    M_1 = \mathbb{E}(s_n \hat{s}_n^*) = 1 - 2Q(\sqrt{2 \gamma}).
\end{align}
Substituting $M_1$ and $M_2$ into the expression of $T_1$, we obtain
\begin{align}
    T_1 = \left\{ 
    \begin{aligned}
        & N + N(N-1) \left[ 1-2 Q \left( \sqrt{2\gamma} \right) \right]^2, \quad k = 0, \\
        & N -N \left[ 1-2 Q \left( \sqrt{2\gamma} \right) \right]^2, \quad k \ne 0.
    \end{aligned}
    \right.    
\end{align}
Since \(|\hat{s}_n|^2 = 1\) for BPSK, the noise-dependent term $T_2$ is given by
\begin{align}
    T_2 = N_0 N \mathbb{E}(|\hat{s}_n|^2) = N_0 N.
\end{align}
Therefore, by substituting the above results into \eqref{ESL 2}, the average mainlobe level and ESL for the BPSK case can be readily obtained as follows
\begin{align} \label{ESL BPSK}
    &\nonumber \mathbb{E}(|r_k|^2)_\mathrm{BPSK} = \\
    &\left\{ 
    \begin{aligned}
        & N + N(N-1) \left[ 1-2 Q \left( \sqrt{2\gamma} \right) \right]^2 + N_0 N, \quad k = 0, \\
        & N -N \left[ 1-2 Q \left( \sqrt{2\gamma} \right) \right]^2 + N_0 N, \quad k \ne 0.
    \end{aligned}
    \right.    
\end{align}

\subsection{QPSK}
We next consider the QPSK case, where each transmitted symbol takes values from the constellation 
\begin{align}
\left\{\frac{1 + j}{\sqrt{2}},\frac{-1+j}{\sqrt{2}},\frac{-1-j}{\sqrt{2}},\frac{1-j}{\sqrt{2}}\right\}.    
\end{align} 
Since all QPSK constellation points have unit magnitude, we directly have
\begin{align}
    M_2 = \mathbb{E}(|s_n|^2 |\hat{s}_n|^2) = 1.
\end{align}
Under AWGN and hard-decision detection, the in-phase and quadrature components are detected independently. Therefore, the probability of correctly detecting a transmitted symbol is
\begin{align}
    \mathrm{Pr}(c_n = 1) = \left[ 1 - Q \left( \sqrt{\gamma} \right) \right]^2.
\end{align}
The probabilities of detecting a transmitted symbol as the two adjacent constellation points are given by
\begin{align}
\begin{aligned}
    & \mathrm{Pr}(c_n = j) = Q \left( \sqrt{\gamma} \right) -  Q^2 \left( \sqrt{\gamma} \right),\\
    & \mathrm{Pr}(c_n = -j) = Q \left( \sqrt{\gamma} \right) -  Q^2 \left( \sqrt{\gamma} \right).
\end{aligned}
\end{align}
 In addition, the probability of detecting the transmitted symbol as the diagonally opposite constellation point is
\begin{align}
    \mathrm{Pr}(c_n = -1) = Q^2 \left( \sqrt{\gamma} \right).
\end{align}
Based on the above decision probabilities, $M_1$ is obtained as
\begin{align}
    M_1 = \mathbb{E}(s_n \hat{s}_n^*) =  1 - 2Q(\sqrt{\gamma}).
\end{align}
Substituting $M_1$ and $M_2$ into the expression of $T_1$ yields
\begin{align}
    T_1 = \left\{ 
    \begin{aligned}
        & N + N(N-1) \left[ 1 - 2 Q \left( \sqrt{\gamma} \right) \right]^2, \quad k = 0, \\
        & N -N \left[ 1 - 2 Q \left( \sqrt{\gamma} \right) \right]^2, \quad k \ne 0.
    \end{aligned}
    \right.    
\end{align}
Since QPSK is also a constant-modulus modulation, the noise-related term $T_2$ of QPSK is
\begin{align}
    T_2 = N_0 N \mathbb{E}(|\hat{s}_n|^2) = N_0 N.
\end{align}
Accordingly, the average mainlobe level and ESL for the QPSK case are
\begin{align} \label{ESL QPSK}
    &\nonumber \mathbb{E}(|r_k|^2)_{\mathrm{QPSK}} = \\
    &\left\{ 
    \begin{aligned}
        & N + N(N-1) \left[ 1 - 2 Q \left( \sqrt{\gamma} \right) \right]^2 + N_0 N, \quad k = 0, \\
        & N -N \left[ 1 - 2 Q \left( \sqrt{\gamma} \right) \right]^2 + N_0 N, \quad k \ne 0.
    \end{aligned}
    \right.    
\end{align}

\subsection{Square QAM}
In this subsection, a general square \(M\)-QAM constellation is considered. First of all, note that each \(M\)-QAM symbol can be decomposed into two independent \(L\)-PAM components as
\begin{align}
    M\text{-QAM} = L\text{-PAM} + j L\text{-PAM}, \quad L^2 = M.
\end{align}
Accordingly, the transmitted and detected symbols can be rewritten as
\begin{align}
    s_n = \alpha_n + j \beta_n, \quad \hat{s}_n = \hat{\alpha}_n + j \hat{\beta}_n,
\end{align}
where $\alpha_n$ and $\beta_n$ denote the in-phase and quadrature $L$-PAM components, respectively.

For the normalized square $M$-QAM constellation, each $L$-PAM component takes values from
\begin{align}
    \alpha_n 
    \in \ell \left\{ m_1, m_2, ..., m_{L-1}, m_L \right\},
\end{align}
where $\ell = \sqrt{\frac{3}{2(L^2 - 1)}}$ and $m_q = -(L-1) + 2(q-1)$. Under AWGN, the noise variance on each real dimension is
\begin{align}
    \sigma^2 = {N_0}/{2} = {1}/{2 \gamma}.
\end{align}
For notational convenience, define
\begin{align}
    \tilde{\gamma} := \frac{\sqrt{\frac{3}{2(L^2 - 1)}}}{\sigma}
    = \frac{\sqrt{\frac{3}{2(L^2 - 1)}}}{\sqrt{\frac{1}{2 \gamma}}}
    = \sqrt{\frac{3 \gamma }{(L^2 - 1)}}.
\end{align}
The parameter $\tilde{\gamma}$ denotes the normalized decision margin, defined as the ratio between the distance from a constellation point to its nearest decision threshold and the noise standard deviation. It quantifies the effective reliability of the decision process at the detection threshold, and therefore governs the symbol transition probabilities and the overall detection performance of $L$-PAM signaling.

By symmetry and the independence between the in-phase and quadrature decision processes, we have
\begin{align}
        M_1 &\nonumber = \mathbb{E}(s_n \hat{s}_n^*)  = \mathbb{E} \left\{ (\alpha_n + j \beta_n) (\hat{\alpha}_n - j \hat{\beta}_n) \right\} \\
        &= \mathbb{E}(\alpha_n \hat{\alpha}_n) + \mathbb{E}(\beta_n \hat{\beta}_n) = 2 \mathbb{E}(\alpha_n \hat{\alpha}_n).
\end{align}
Similarly, $M_2$ is given by
\begin{align}
    &\nonumber M_2  = \mathbb{E}(|s_n|^2 |\hat{s}_n|^2)  = \mathbb{E} \left\{ (|\alpha_n|^2 + |\beta_n|^2) (|\hat{\alpha}_n|^2 + |\hat{\beta}_n|^2) \right \}\\
   &\nonumber = \mathbb{E} \left \{  |\alpha_n|^2 |\hat{\alpha}_n|^2 + |\alpha_n|^2 |\hat{\beta}_n|^2 + |\beta_n|^2 |\hat{\alpha}_n|^2 + |\beta_n|^2 |\hat{\beta}_n|^2\right \} \\
    & = 2 \mathbb{E} \left( |\alpha_n\hat{\alpha}_n|^2 \right) + 2 \mathbb{E}(|\alpha_n|^2) \mathbb{E} (|\hat{\alpha}_n|^2).
\end{align}
It can be seen that the evaluation of both $M_1$ and $M_2$ reduces to the one-dimensional $L$-PAM decision statistics. Therefore, we first characterize the conditional decision probability
\begin{align}
    \text{Pr} \left( \hat{\alpha}_n = {\alpha}_\theta | {\alpha}_n = {\alpha}_q \right)  = \text{Pr} \left( \hat{\alpha}_n = \ell m_\theta | {\alpha}_n = \ell m_q \right),
\end{align}
where $\theta, q = 1,2,...,L$. Under AWGN and hard-decision detection, the above conditional probability is determined by the corresponding decision region of each \(L\)-PAM point. Specifically, the conditional decision probabilities can be characterized by the following three cases.
\begin{align}
    &\nonumber \text{Pr} \left( \hat{\alpha}_n = \ell m_\theta | {\alpha}_n = \ell m_q \right) = \\
    &\left\{
    \begin{aligned}
        & 1 - Q\left( (m_1 + 1 - m_q) \tilde{\gamma} \right), \theta=1 \\
        & Q((m_\theta - 1 - m_q) \tilde{\gamma}) - Q((m_\theta + 1 - m_q) \tilde{\gamma}), 2 \leq \theta \leq L-1\\
        &Q((m_L - 1 - m_q) \tilde{\gamma}), \theta=L
    \end{aligned}\right.
\end{align}

Based on the above conditional probabilities, the required one-dimensional moments can be derived accordingly. Since the $L$-PAM symbols are equiprobable, we first obtain
\begin{align}
    \mathbb{E}(|\alpha_n|^2) = \frac{\ell ^2}{L} \sum_{q=1}^L |m_q|^2 =\frac{1}{2}.
\end{align}
The second-order moment of the detected in-phase component can be written as
\begin{align} \label{Part_1}
&\nonumber \mathbb{E}\big(|\hat{\alpha}_n|^2\big) 
= \frac{\ell^2}{L} \sum_{q=1}^{L} \\
&\left\{
\begin{aligned}
& m_1^2 \Big[ 1 - Q((m_1 + 1 - m_q)\tilde{\gamma}) \Big] + \\
&\sum_{\theta=2}^{L-1} m_\theta^2 
\Big[
Q((m_\theta - 1 - m_q)\tilde{\gamma})
-
Q((m_\theta + 1 - m_q)\tilde{\gamma})
\Big] \\
& + m_L^2 Q((m_L - 1 - m_q)\tilde{\gamma})
\end{aligned}
\right\}.
\end{align}
Similarly, the correlation term $\mathbb{E}(\alpha_n\hat{\alpha}_n)$ is given by 
\begin{align} \label{Part_2}
&\nonumber \mathbb{E}\big(\alpha_n \hat{\alpha}_n\big)
= \frac{\ell^2}{L} \sum_{q=1}^{L}\\
&\left\{
\begin{aligned}
& m_q m_1 \Big[ 1 - Q((m_1 + 1 - m_q)\tilde{\gamma}) \Big] + \\
& \sum_{\theta=2}^{L-1} m_q m_\theta
\Big[
Q((m_\theta - 1 - m_q)\tilde{\gamma})
-
Q((m_\theta + 1 - m_q)\tilde{\gamma})
\Big] \\
& + m_q m_L\, Q((m_L - 1 - m_q)\tilde{\gamma})
\end{aligned}
\right\}.
\end{align}
Moreover, the fourth-order mixed moment can be expressed as \begin{align}\label{Part_3}
&\nonumber \mathbb{E}\!\left(|\alpha_n \hat{\alpha}_n|^2\right)
= \frac{\ell ^4}{L}\sum_{q=1}^{L} \\
&\left\{
\begin{aligned}
& m_q^2 m_1^2 \Big[ 1 - Q((m_1 + 1 - m_q)\tilde{\gamma}) \Big] +\\
& \sum_{\theta=2}^{L-1} m_q^2 m_\theta^2
\Big[ Q((m_\theta - 1 - m_q)\tilde{\gamma}) - Q((m_\theta + 1 - m_q)\tilde{\gamma}) \Big] \\
& + m_q^2 m_L^2 \, Q((m_L - 1 - m_q)\tilde{\gamma})
\end{aligned}
\right\}.
\end{align}
Accordingly, $M_1$ and $M_2$ for square $M$-QAM can be obtained by substituting \(\mathbb{E}(|\alpha_n|^2)\), \(\mathbb{E}(|\hat{\alpha}_n|^2)\), \(\mathbb{E}(\alpha_n\hat{\alpha}_n)\), and \(\mathbb{E}(|\alpha_n\hat{\alpha}_n|^2)\) into the previously derived expressions. Moreover, from \eqref{T2} and the structure of \(\hat{s}_n\), the noise-dependent term is given by
\begin{align}
    T_2 = N_0 N \mathbb{E}(|\hat{s}_n|^2) = 2 N_0 N \mathbb{E}(|\hat{\alpha}_n|^2).
\end{align}
Therefore, once the closed-form expressions of $M_1$, $M_2$, and $T_2$ are obtained, the average mainlobe level and the ESL for square $M$-QAM follow directly from \eqref{ESL 2}.

\section{Simulation}
\begin{figure*}[t]
    \centering
    \subfigure[BPSK]{
        \includegraphics[width=0.68\columnwidth]{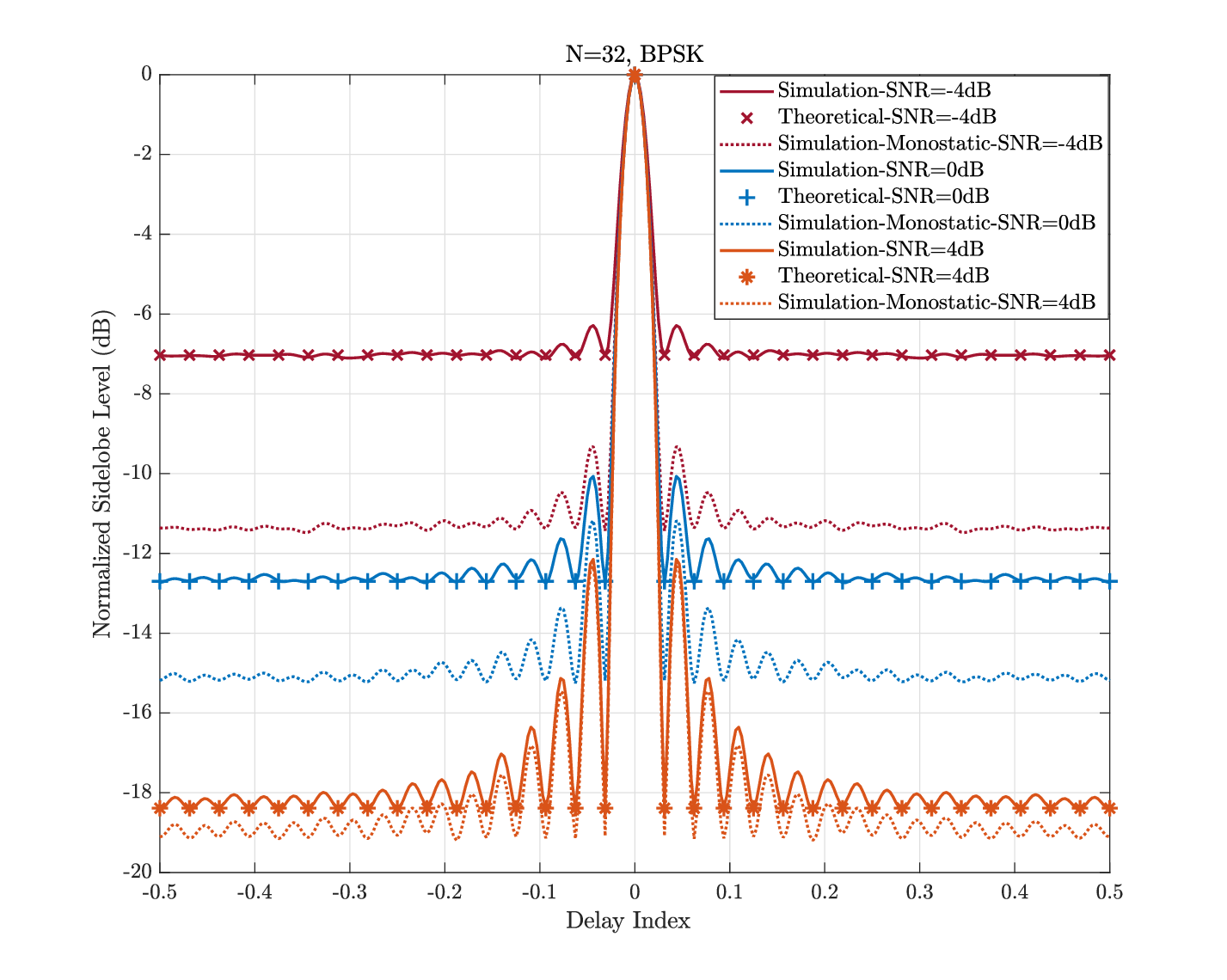}
        \label{fig: range profile of BPSK}    
    } \hspace{-7mm}
    \subfigure[QPSK]{
        \includegraphics[width=0.68\columnwidth]{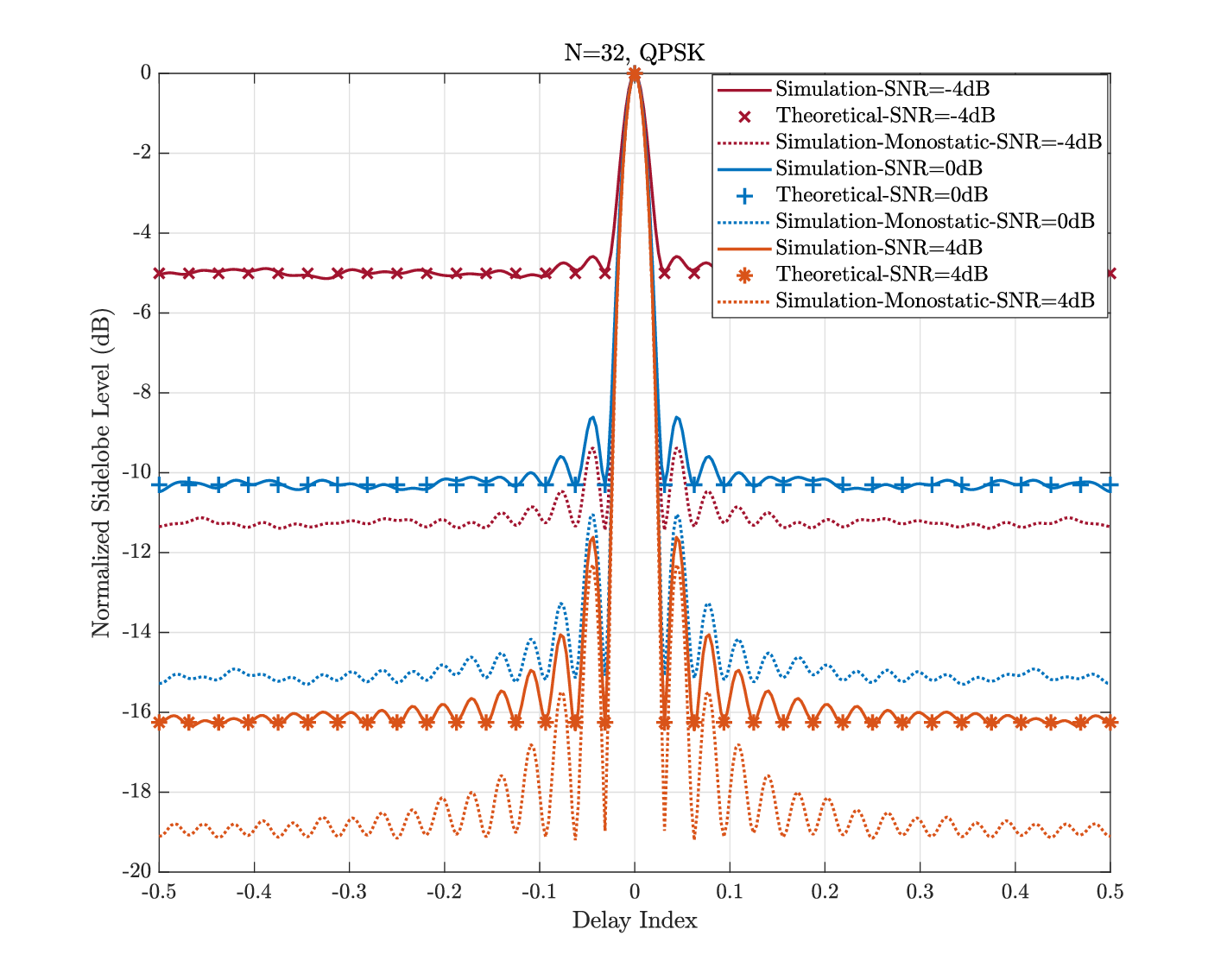}
        \label{fig: range profile of QPSK}    
    } \hspace{-7mm}
    \subfigure[64-QAM]{
        \includegraphics[width=0.68\columnwidth]{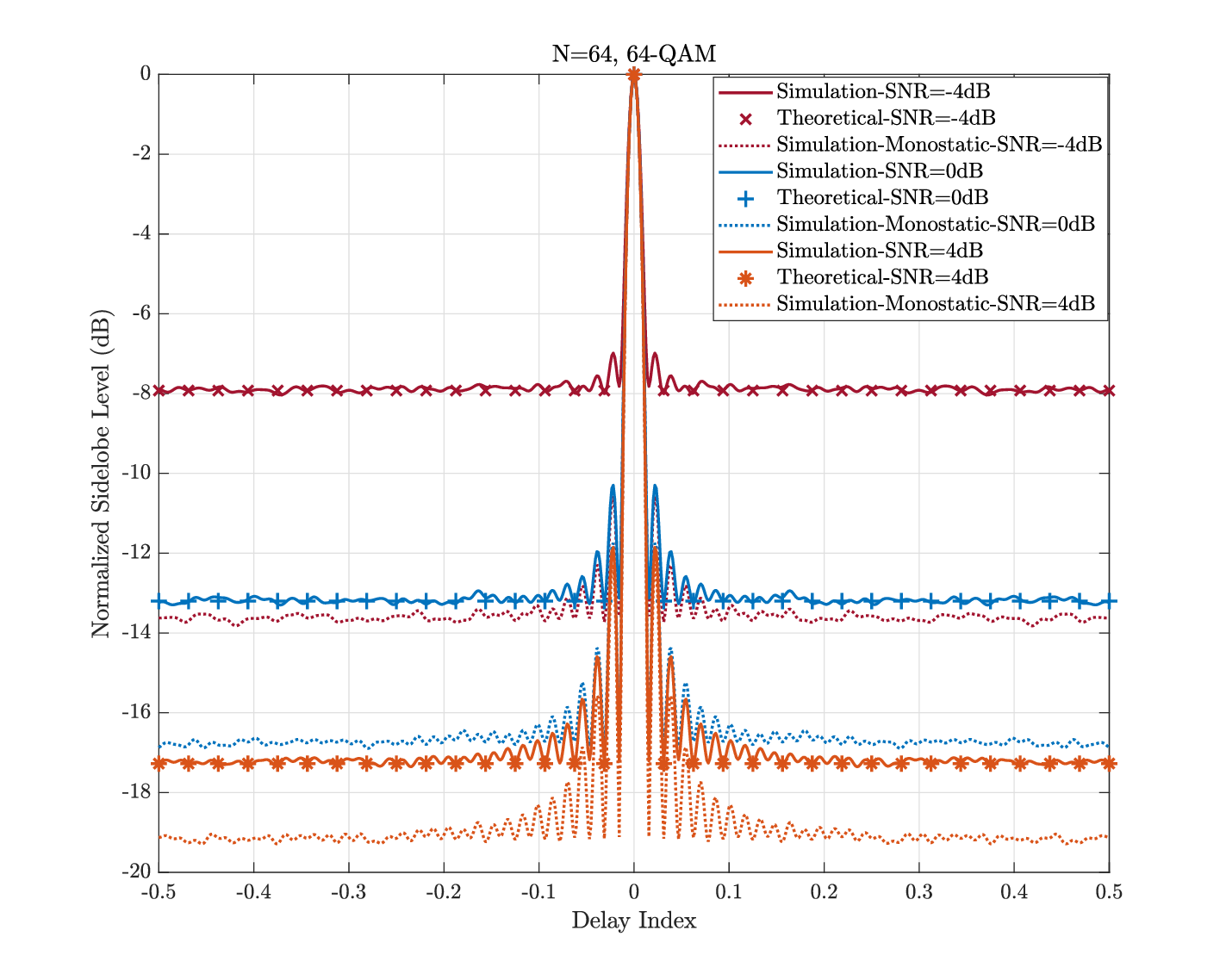}
        \label{fig: range profile of QAM64}    
    }
    \caption{The range profiles of BPSK, QPSK and 64-QAM.} 
    \label{fig: range profile}
\end{figure*}

\begin{figure}[h]
    \centering
    \includegraphics[width=0.8\linewidth]{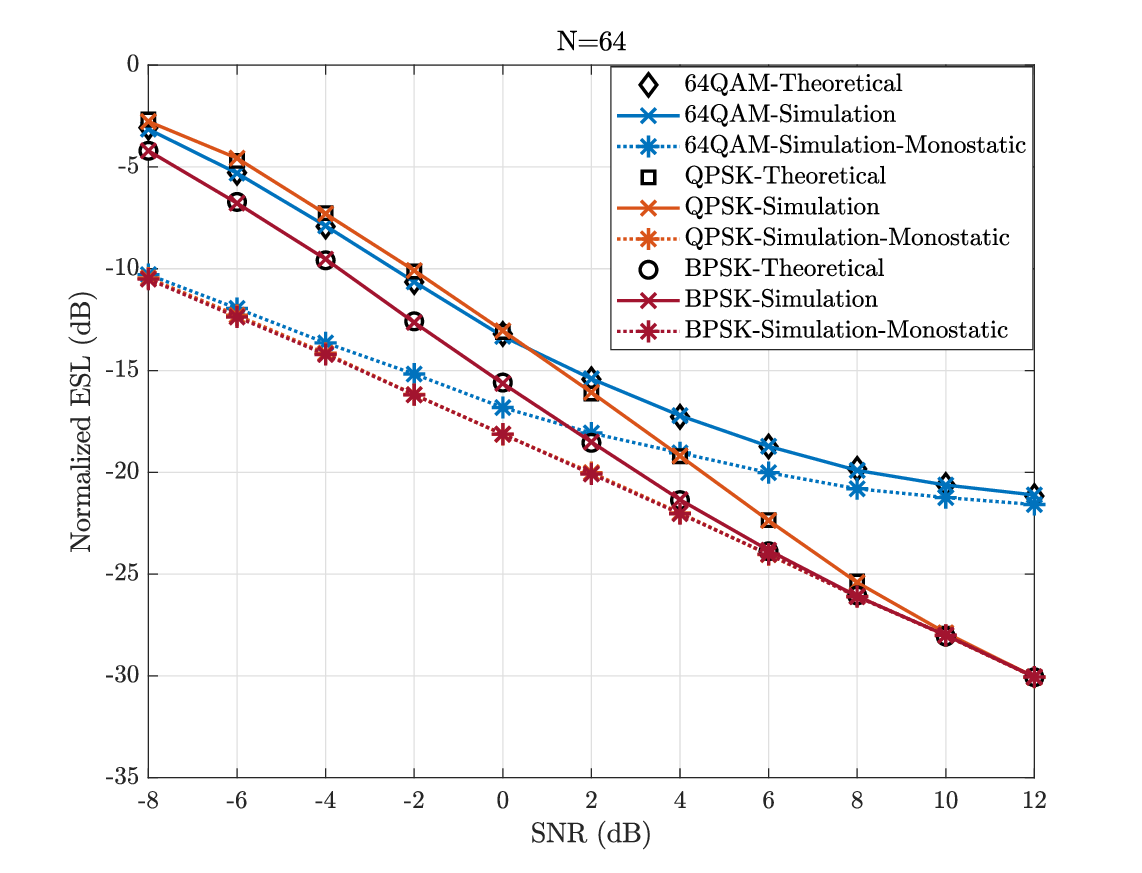}
    \vspace{-5mm}
    \caption{The normalized ESL performances of BPSK, QPSK and 64-QAM.}
    \label{fig: Normalized ESL}
\end{figure}

In this section, numerical results are presented to further validate our findings and theoretical analysis. All simulation results are obtained by averaging over \(1000\) independent random realizations. Fig.~\ref{fig: range profile} shows the range profiles for BPSK, QPSK, and 64-QAM, respectively, while Fig.~\ref{fig: Normalized ESL} compares the normalized ESL performances of BPSK, QPSK, and square QAM under different SNR values. It should be noted that the range-profile results are generated with \(10\times\) oversampling in order to provide a finer visualization of the sidelobe structure.

Several observations can be made from these results. First, for all considered modulation schemes, the theoretical and simulation results are in close agreement, thereby validating the derived closed-form expressions. Second, the ideal matched-filtering result, where the transmitted signal is directly used as the reference for correlation, is included for each SNR as a benchmark and is labeled as ``Simulation-Monostatic'' in the figures. Comparison with this benchmark shows that demodulation errors degrade the sidelobe performance. Specifically, using the demodulated symbols as the reference leads to a higher normalized sidelobe level than that of the ideal matched-filtering case.

Moreover, as the SNR increases, the gap between the normalized sidelobe level achieved by mismatched filtering and that achieved by ideal matched filtering becomes progressively smaller. This observation indicates that the mismatch between the reconstructed reference and the true transmitted symbol sequence is gradually reduced at higher SNR.

\section{Conclusion}
In this work, we investigated the ranging performance of a bistatic OFDM-based ISAC system, in which demodulated symbols serve as the sensing reference. The results indicate that symbol decision errors induce mismatched filtering, making the delay-domain correlation response explicitly dependent on the SNR. For the frequency-flat fading case, closed-form expressions for the ESL and average mainlobe level were derived under BPSK, QPSK, and general square QAM constellations. The theoretical analysis was further validated through numerical simulations. Future work may extend the study to frequency-selective channels.

\balance
\bibliographystyle{IEEEtran}
\bibliography{refs}
\end{document}